\documentclass{aa520}
\usepackage{graphicx}
\def\lya{Ly$\alpha$~}
\begin{document}
   \title{A VLT/FORS2 spectroscopic survey in the HDF-S 
\thanks{Based on observations
collected at the European Southern Observatory, Paranal, Chile}  
}
   \author{Vanzella E.\inst{1,2}
        \and
        Cristiani S.\inst{3,4}
        \and
        Arnouts S.\inst{1}
        \and
        Dennefeld M.\inst{5}
        \and
        Fontana A.\inst{6} 
        \and
        Grazian A. \inst{1,2}
        \and
        Nonino M. \inst{3}
        \and
        Petitjean P. \inst{5}
        \and
        Saracco P. \inst{7}
        }
   \offprints{E.Vanzella (evanzell@eso.org)}
   \institute{European Southern Observatory,
        Karl-Schwarzschild-Str. 2, 
        D-85748 Garching, Germany
   \and
        Dipartimento di Astronomia dell'Universit\`a di Padova, 
        Vicolo dell'Osservatorio 2, 
        I-35122 Padova, Italy
   \and 
        INAF -Osservatorio Astronomico di Trieste, Via G.B. Tiepolo 11, 
        40131 Trieste, Italy
   \and
        Space Telescope European Coordinating Facility,
        Karl-Schwarzschild-Str. 2, D-85748 Garching, 
        Germany
   \and
        Institut d'Astrophysique de Paris, 98\,{\it bis},
Boulevard Arago, F-75014 Paris, France 
   \and 
        INAF - Osservatorio Astronomico di Roma, via
        dell'Osservatorio 2, Monteporzio, Italy
   \and
        INAF - Osservatorio Astronomico di Brera, via E. Bianchi 46,
        Merate, Italy
} 
   \date{Received \dots; accepted \dots}

   \abstract{
    We report on low-resolution multi-object spectroscopy of 65
    objects from $I(AB) \simeq 20$ to $I(AB) \simeq 25$
    in the HDF-S obtained with the VLT
    Focal Reducer/low dispersion Spectrograph (FORS2). 
    18 objects belong to the HDF-S proper, i.e. the WFPC2 deep area. 
    15 high-redshift galaxies with $2.0
    < z < 3.5$  (10 in the HDF-S proper) have been identified.
    The spectroscopic redshifts are in good agreement with the
    photometric ones derived from a $\chi^{2}$ minimization
    technique comparing the observed spectral energy distribution with
    synthetic libraries and with a new neural network ($NN$)
    approach. The dispersion with the former method is $\sigma_z=0.16$
    whereas the latter provides $\sigma_z=0.13$.  No
    ``catastrophic'' difference is encountered.
    The inferred star formation rates of the individual objects range 
    from tens to a few hundreds of $M_{\odot}$ yr$^{-1}$ and the global
    star formation rate of the Universe at $<z>\simeq 2.4$ is estimated to be 
    0.15  $M_{\odot}$ yr$^{-1}$ $Mpc^{-3}$ with a statistical error of 0.04. 

    Evidence for large scale structure is found with two groups' redshifts
    observed at $z\simeq 2.1$ and $z\simeq 2.7$ and a pronounced
    low redshift peak around $z\simeq 0.58$.

    An elliptical galaxy lensing a background object turns out to be
    at a redshift $z=0.577$. 

} 
   \maketitle
%
%________________________________________________________________

\section{Introduction}

The Hubble Deep Field South (HDF-S) is a region of intense
astrophysical interest.
A large set of observations of an otherwise unremarkable field around
the QSO J2233-606 (z $=$ 2.24) has been taken in parallel by three
instruments aboard the Hubble Space Telescope (HST): the Wide Field
and Planetary Camera 2 (WFPC2), the Space Telescope Imaging
Spectrograph (STIS) and the Near Infrared Camera and Multi-Object Spectrometer (NICMOS).
This unique database has been complemented with a vigorous campaign of
ground-based observations from the optical to the radio
by virtually all the major observatories in the southern hemisphere.

In a series of papers we have presented deep near-IR observations of
the HDF-S (\cite{saracco01}), a multicolor ($U, B, V, I, Js, H, Ks$)
catalog in the WFPC2 area (\cite{vanzella01}), preliminary
spectroscopic identifications (\cite{cristiani00a}), photometric
redshifts (\cite{fontana02}; Vanzella et al. in preparation), studies
of the redshift distribution and evolution of the galaxy luminosity
function (\cite{fontana99,fontana00,poli01}) and clustering
(\cite{arnouts02}), as well as high-resolution spectroscopy of the
HDF-S QSO J2233-606 (\cite{cristiani00b}).

In the present paper we report the results of a spectroscopic campaign
aiming at the confirmation of all the high-redshift galaxy candidates
in the WFPC2 area brighter than $I(AB)=24.25$.  A
cosmology with $H_o=70 Km~s^{-1}Mpc^{-1}$, $\Omega_{M}=0.3$ and
$\Omega_{\Lambda}=0.7$ is assumed throughout.

%__________________________________________________________________

\section{The photometric databases and the selection of the
candidates} Deep multicolor imaging of the HDF-S has been obtained
from the space and from the ground. In particular the WFPC2 data,
consisting of deep images in the F300W, F450W, F606W and F814W
filters, cover an area of 4.7 sq.arcmin reaching $10 \sigma$ AB magnitude
limits of 26.8, 27.7, 28.2 and 27.7 (in a 0.2 sq.arcsec area).
$UBVRIJHK$ data over an area of 25 sq.arcmin, including the WFPC2
field, have been obtained at the ESO 3.5m New Technology Telescope
(NTT) as a part of the ESO Imaging Survey (EIS) program (\cite{EISHDFS}).
They reach  $2\sigma$ limiting magnitudes of
$U_{AB}\sim 27$, $B_{AB}\sim 26.5$, $V_{AB}\sim 26$, $R_{AB}\sim 26$,
$I_{AB}\sim 25$, $J_{AB}\sim 25$, $H_{AB}\sim 24$ and $K_{AB}\sim
24$, in 2xFWHM diameter apertures.
Near infrared images centered in the WFPC2 field have been obtained with the
ISAAC instrument at the VLT in the filters Js, H and Ks. 
A first set of observations, obtained in 1999, have been published
by Saracco et al. (2001) and reach the AB-magnitude 
24.4, 23.5, 24.0 in the Js, H and Ks band respectively, at $5 \sigma$  
in a 2$\times$FWHM diameter aperture.
The photometric catalog is described in (\cite{vanzella01}).
The final set with the complete reduction is currently being analyzed
and reaches 25.0, 24.0, 24.5 in the Js, H and Ks band, respectively.

\begin{figure}
\includegraphics[width=90mm]{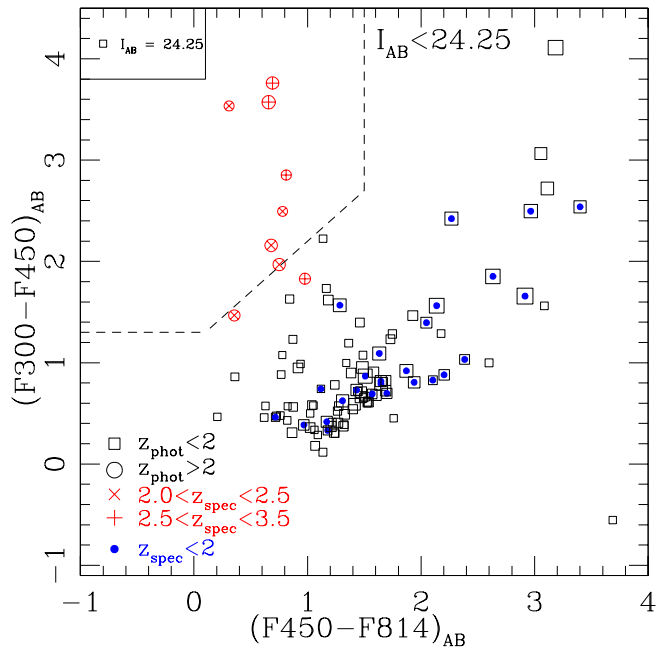}
\caption{Selection of the high redshift galaxies ($U$ band $dropout$). 
The dashed line outlines the two-color selection criterion. 
The size of the symbols is inversely proportional to the magnitude
of the objects.}
\label{fig:zsel}
\end{figure}

%__________________________________________________________________

\section{Spectroscopic Observations}

\begin{table}
\caption{Journal of the MOS Observations}
\begin{tabular}{lcccc}
\hline \hline
Field & $\alpha_{2000}$ & $\delta_{2000}$ & date &
exp.time (ks) \\
\hline
A   & 22:32:54 &$-60$:33:54 & 2000-Sep-1 & $4 \times 2.7 + 2.4$ \\
\\
B   & 22:32:51 &$-60$:34:14 & 2000-Sep-2 & $4 \times 3.6$ \\
\\
D   & 22:32:54 &$-60$:33:54 & 2000-Sep-2 & $4 \times 3.6 +1.2$ \\
\hline
\hline
\label{tab:obs}
\end{tabular}
\end{table}

The main emphasis of the present work is on high-redshift galaxies.

In the HDF-S proper, i.e. in the WFPC2 deep area, 
high redshift  ($z \geq 2$) galaxy candidates were selected on
the basis of two-color diagrams F300-F450 vs. F450-F814, 
see Fig.~\ref{fig:zsel}). 
Spectroscopic observations of the nine high-z galaxy candidates
with $I(AB) < 24.25$ have been carried out, confirming
all of them to be galaxies with $2<z<3.5$.

Outside the WFPC2 field, targets were selected from a list of
Lyman-break candidates produced by the EIS project (astro-ph/9812105).

The present spectroscopic observations were carried out with the FORS2
instrument (\cite{FORS2}) in multiple object spectroscopy (MOS) mode
on September 2000.

In the FORS2 MOS mode, 19 pairs of movable slit blades can be placed in a
FOV of $6.8 \times 6.8$ sq.arcmin. The actual useful field in the
direction of the dispersion is somewhat smaller and depends on the length
of the spectra/dispersion. In the present case the grism I150 was
used, providing a useful field of $3.5 \times 6.8$ sq.arcmin.
The journal of the observations is given in Table~\ref{tab:obs}.
Three different pointings were observed.

In the choice of the objects to be observed priority was given to high
redshift galaxy candidates with $I(AB)<24.25$. 
When no suitable candidate was available for the allowed range of
positions of a given slit, a random object in the field was chosen.
In a number of cases more than one object has been placed in a given
slit and/or the position of the slit has been changed from one
exposure to another of a given pointing.  In this way the final number
of obtained spectra exceeds 19 per pointing.

\begin{figure*}
\centering
\includegraphics[width=170mm]{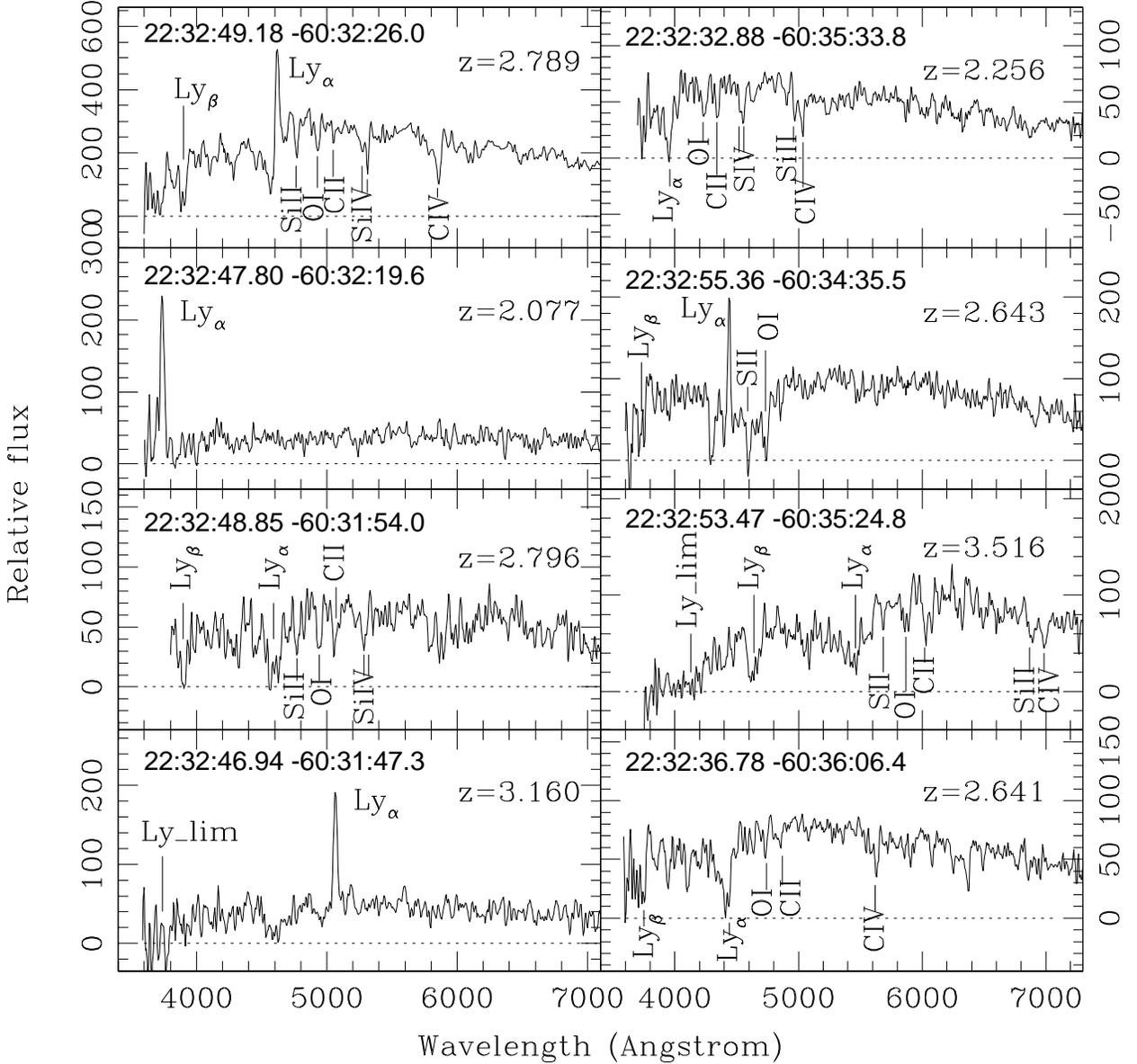}
\caption{Spectra of 8 high-redshift galaxies observed in the
HDF-S. The ordinate gives the relative flux density per Angstrom. The identification numbers in 
the upper left of the panels provide the RA and Declination for each target.}
\label{fig:spectraA}
\end{figure*}

\begin{figure*}
\centering
\includegraphics[width=170mm]{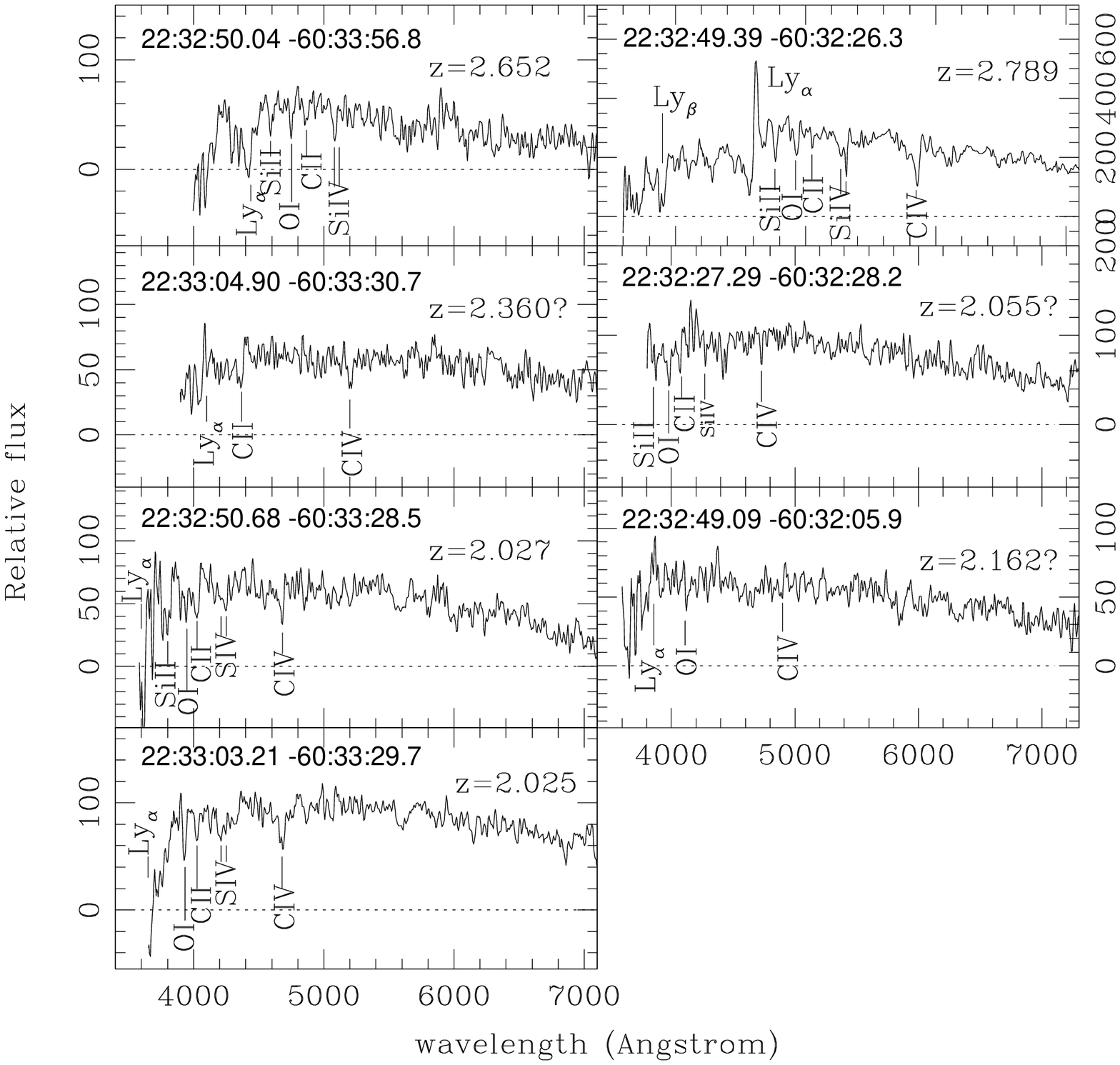}
\caption{Spectra of 7 high-redshift galaxies observed in the HDF-S. The ordinate 
gives the relative flux density per Angstrom. The identification numbers in 
the upper left of the panels provide the RA and Declination for each target.}
\label{fig:spectraB}
\end{figure*}

The MOS observations were reduced with the MIDAS package, using
commands of the LONG and MOS contexts. For each object the available
2-D spectra were stacked and then an optimal extraction was carried
out.
The resulting spectroscopic identifications are listed in Table~\ref{tab:HDFS}
and the spectra of the 15 galaxies with redshifts larger than 2
are shown in Fig.~\ref{fig:spectraA} and Fig.~\ref{fig:spectraB}. 

The photometric data of Table~\ref{tab:HDFS} Column 5 
have been taken from the multicolor
catalog of the HDF-S produced by \cite{vanzella01}.

\begin{figure}[h]
\includegraphics[width=90mm]{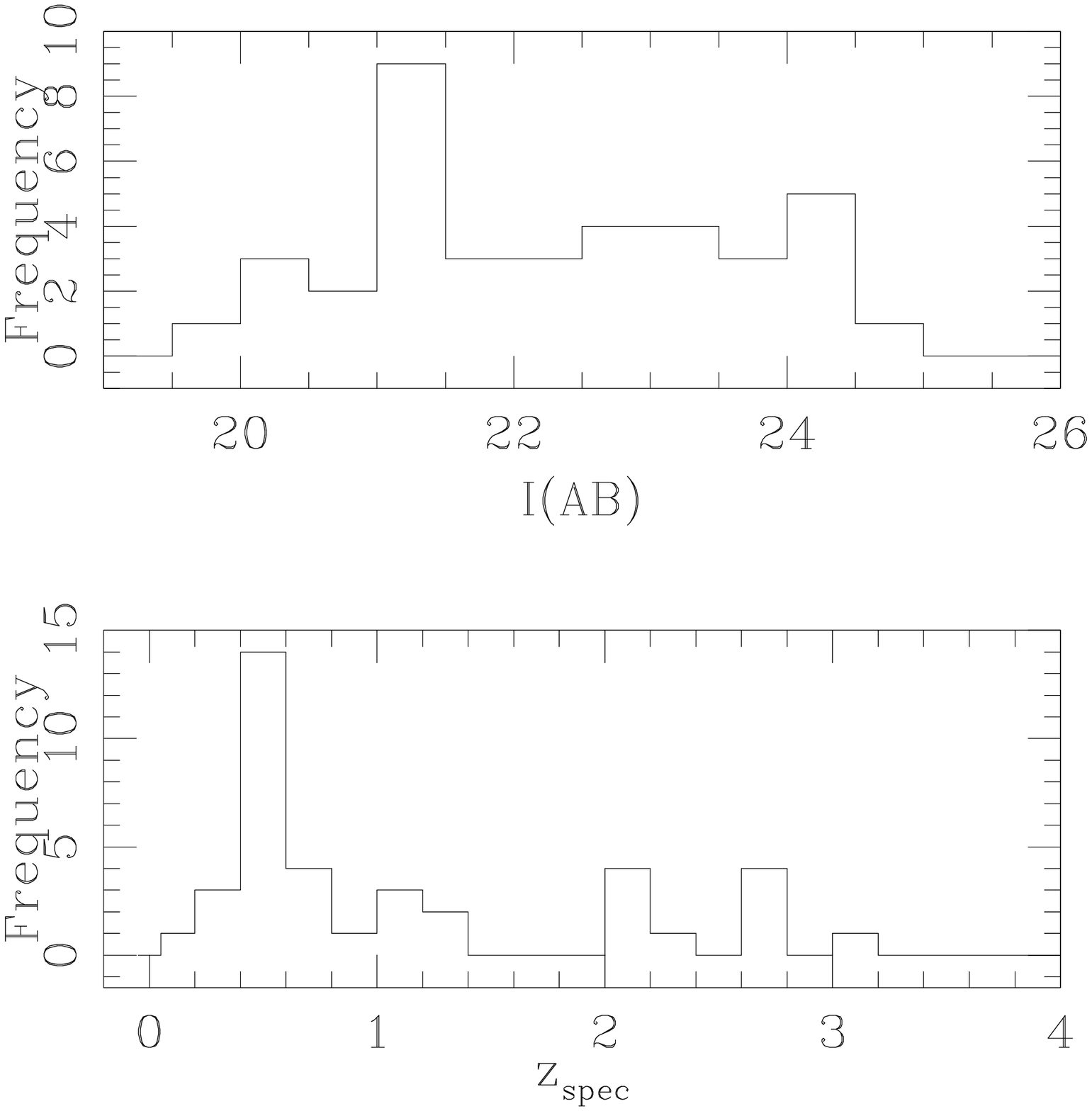}
\caption{In the upper panel the magnitude distribution is shown. In the lower panel 
the redshift distribution is shown. In both cases for the spectroscopic sample 
in the WFPC2, including the previous observations (see text).}
\label{fig:z_mag_distr}
\end{figure}

Fig.~\ref{fig:z_mag_distr} shows the magnitude distribution (upper panel) and the
redshift distribution (lower panel) for the spectroscopic sample in the WFPC2, 
including previous observation in the literature (Glazebrook et al. 
{\tt http://www.aao.gov.au/hdfs/Redshifts/}, \cite{cristiani00a},
\cite{rigopo00}, \cite{franc02}). It is worth to note that the spectroscopic sample is $100\%$ complete down to 
$I_{AB} \simeq 21$, and $75\%$ down to $I_{AB} \simeq 22.5$.

The behavior of the redshift distribution shows an evident peak at 
redshift $z \simeq 0.58$, indicating the presence of large scale 
structure, (see also, \cite{Denn01}). More details will be discussed in a forthcoming paper 
(Dennefeld et al. in preparation).

\section{Reliability of the photometric redshift}

The present observations, together with previously published data
(Glazebrook et al. {\tt http://www.aao.gov.au/hdfs/Redshifts/}, 
\cite{cristiani00a}, \cite{rigopo00}, \cite{franc02}) provide
46 spectroscopic identifications in the WFPC2 area, 
the full list is given in Table~\ref{tab:zspecALL}.
Eight objects, marked with a colon (:) in Col. 1, have unreliable 
photometry (they are near the border of the WF camera).

\begin{figure}
\includegraphics[width=90mm]{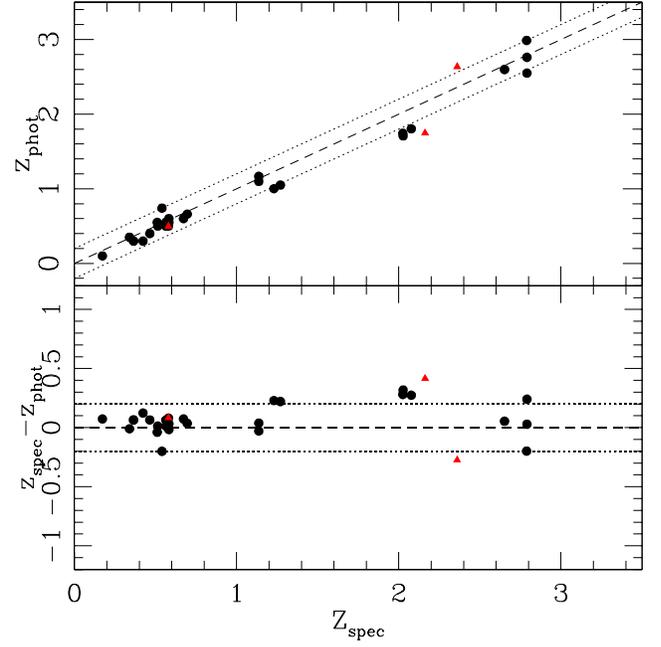}
\caption{Predictions with the $\chi^{2}$ minimization technique in the
HDF-S spectroscopic sample. Objects with
the unreliable photometry have not been considered. The triangles represent the uncertain
spectroscopic redshifts.}
\label{fig:BC99font}
\end{figure}

Photometric redshifts in the WFPC2 field have been computed adopting
two techniques.  The first method is based on a $\chi^{2}$
minimization comparing the observed magnitudes (SED) with synthetic
libraries (\cite{arnouts02}, \cite{fontana00}). 
Fig.~\ref{fig:BC99font} shows the comparison between
photometric and spectroscopic redshifts for the sample available in
the WFPC2 field using the GISSEL00 version of the package by Bruzual \& Charlot
(1993). Objects with unreliable photometry have not been considered.
More details will be discussed in a companion paper (\cite{fontana02}).

\begin{figure}
\includegraphics[width=90mm]{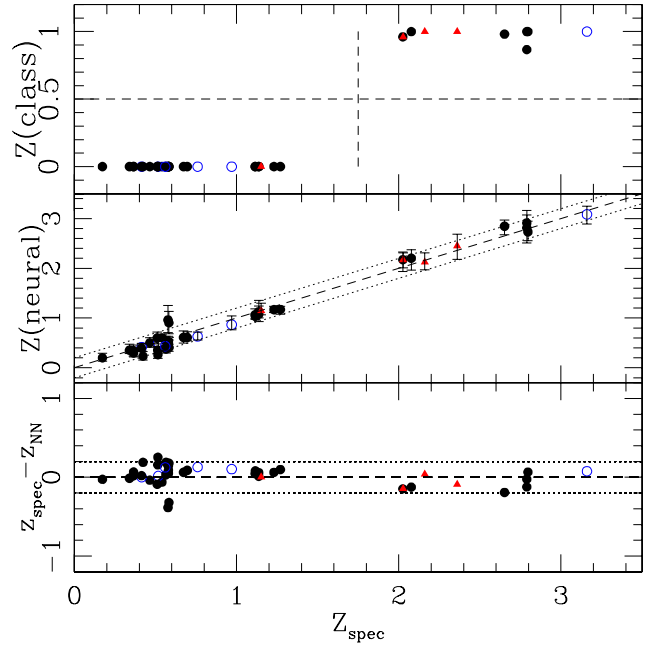}
\caption{Predictions with the neural network approach in the HDF-S
spectroscopic sample. In the upper panel the classification between
low and high redshift galaxies, less than 1.75 and greater than 1.75,
respectively. In the middle and lower panels the comparison between
neural photometric redshifts and the spectroscopic ones. Open circles
represent objects with unreliable photometry and triangles are objects
with uncertain spectroscopic redshift.
}
\label{fig:DZneural}
\end{figure}

The second method is based on a neural network ($NN$) approach
(Vanzella et al. in preparation).  Fig.~\ref{fig:DZneural} shows the
results of the prediction on the spectroscopic sample of the HDF-S.
The network used is the classical $multilayer~perceptron$  (MLP) with
the standard $backpropagation$ learning algorithm (for an introduction
to artificial neural networks see C.A.L. Bailer-Jones et
al. astro-ph/0102224).  

The $backpropagation$ has been used to train the MLP on the available spectroscopic 
sample in the HDF-N (\cite{cohen00}) mixed with a set of SEDs computed from the templates of Coleman, Wu \& Weedman (1980).

After a suitable training, based on a bootstrap technique, the $NNs$ have
been applied to the HDF-S spectroscopic sample.

A $NN$ with different architecture has been used as a binary classifier to divide the sources into the 
two classes of high ($z>1.75$) 
and low ($z<1.75$) redshift, corresponding to 1. and 0. respectively (upper panel of 
Fig.~\ref{fig:DZneural}). The classification is correct for all the spectroscopically identified 
objects.
The input pattern in this case contains the colors, the apparent luminosity in the $I$ band 
and the isophotal area of the object (obtained on the basis of the SExtractor package, \cite{bertin96}).

The middle panel of Fig.~\ref{fig:DZneural} shows the direct evaluation of the redshift for the same sample. 
In this case the input pattern consists of the colors.
 
The two methods have produced a similar dispersion, $\sigma_{z}\simeq0.13$ for
the neural approach and $\sigma_{z}\simeq0.16$ for the $\chi^{2}$ minimization technique,
in a sample where the objects with unreliable photometry are not included (38 galaxies).
Including also these objects the neural approach gives again $\sigma_{z}\simeq0.13$.

The general agreement of the photometric redshifts with the
observed ones is remarkable. It should be noted, however, that the 
high-redshift galaxy candidates have been selected on the basis of 
color criteria, which are not entirely independent of the photometric 
redshift approach. For example, if there are galaxies truly at high 
redshift with peculiar colors that the photometric redshift technique 
would predict to be at low-z, they might not have been selected with 
the color criteria described in Sect.2, biasing the measured scatter
in redshift.

\section{High Redshift Galaxies}

In this spectroscopic survey 15 high redshift galaxies ($z>2$) have been identified
(9 are inside the WFPC2 area and one is near its border).

A $F300-F450$ vs. $F450-F814$ diagram of the objects in the WFPC2 area with
$I(AB)<24.25$ is shown in Fig.~\ref{fig:zsel}. Such a plot is
customarily used to find high redshift galaxies and in the present
case 7 galaxies with $z>2$ have been selected according to a standard
color criterion (\cite{madau96}). The photometric redshift techniques
make it possible to increase the completeness of the sample, leading
to the selection of two more galaxies with $z>2$.

\begin{figure}
\includegraphics[width=90mm]{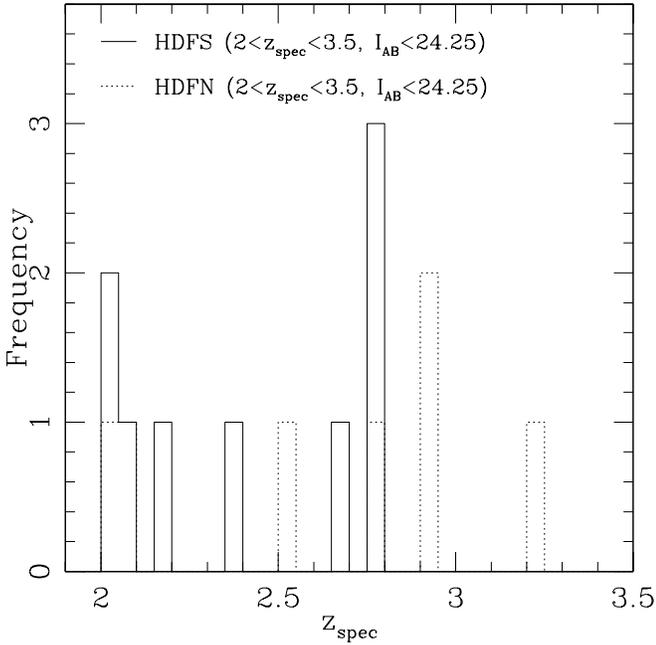}
\caption{Comparison between high redshift galaxies distributions in
the HDF-S (solid line) and HDF-N (dotted line) with at $I_{AB}<24.25$
and $2<z<3.5$.}
\label{fig:hist_hz}
\end{figure}

The redshift distributions of high redshift galaxies ($2<z<3.5$) in
the HDF-N and HDF-S with $I(AB)<24.25$ are shown in
Fig.~\ref{fig:hist_hz}.  Two groups of redshift are apparent in the
HDF-S around z$\simeq2.1$ (4 galaxies) and z$\simeq2.7$ (4
galaxies). The KS test excludes at the 2$\%$ level the hypothesis 
that (neglecting any selection function)
the distribution of redshift in the HDF-S is uniform between $z=2$ and
$z=3.5$.

The comparison with the HDF-N spectroscopic sample at the same magnitude limit gives a KS 
probability less than 10$\%$ that the two samples are drawn from the same parent
population.

It is interesting to note that the galaxy HDF-S 1432 ($z=2.077$) has the same redshift 
as a metal absorption system at
$z=2.07728$ identified by Cristiani \& D'Odorico (2000) in
the spectrum of the HDF-S QSO J2233-606. The system shows at least two 
components and the CIV $\lambda 1548-1550$ doublet, SiII $\lambda
1260$, SiIII $\lambda 1206$ and FeII $\lambda 2382$ transitions are
observed in addition to a strong $\log (N_{HI}) > 15$ \lya.
Coincidences of absorption systems on scales up to several arcminutes
(\cite{vale02}) are observed in the spectra of quasars as the
signature of spatial correlation of high density peaks.
In the present case the distance between the galaxy HDF-S 1432
and the absorber, 6.2 arcmin, corresponds to 3.1 proper
Mpc and is suggestive of a large scale structure at $z \sim 2$, which
would be extremely interesting to further investigate by carrying out
the identification of Lyman-break galaxies in correspondence of the
HDF-S STIS field.

A bright pair ($I(AB)=23.28, 22.92$ and $ID=1277, 1242$ respectively) of
interacting galaxies has been identified at redshift 2.789. The
angular separation is $1.76''$, corresponding to a proper separation of
13.8 Kpc.

\begin{figure}
\includegraphics[width=90mm]{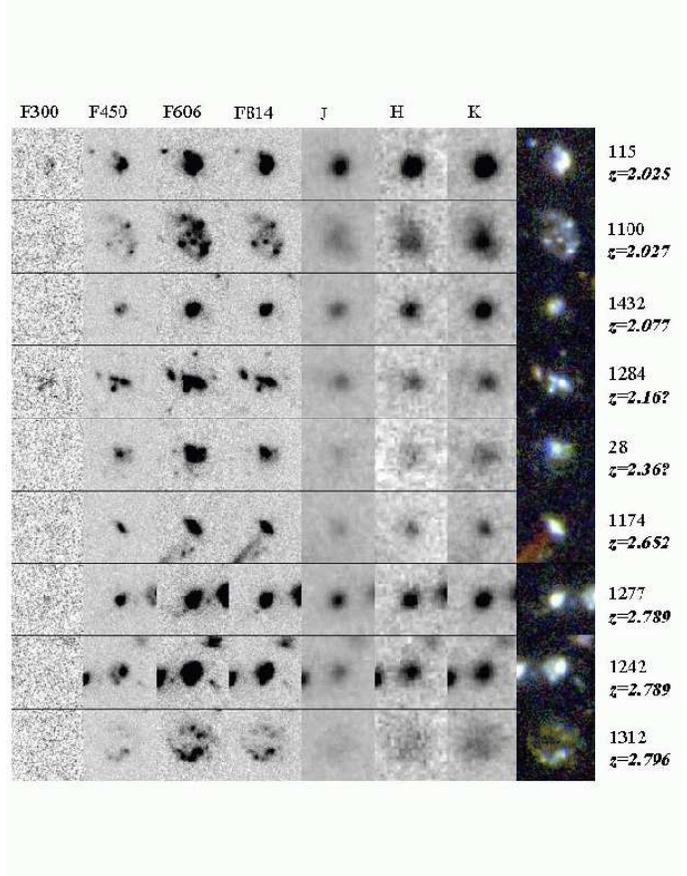} 
\caption{High redshift galaxies in the seven bands and color composition. The side of each box is 3.5 arcsec.}

\label{fig:zmorph}
\end{figure}

Fig.~\ref{fig:zmorph} shows the morphology of each high redshift galaxy in the optical
(WFPC2) and near infrared bands. 

The properties of the high-z galaxies confirmed with the present
observations are summarized in Table~\ref{tab:physics}.
They have been inferred from the GISSEL models best-fitting the
photometric data (\cite{arnouts99}), imposing the redshift at the
spectroscopic value and assuming a continuous star formation.
Col.~4, 5, 6 and 7 list the estimated star formation rate (SFR), age,
stellar mass and $E_{B-V}$ extinction, respectively.

Column~8 lists the SFR empirically derived from the photometry 
and corrected for the intrinsic absorption
according to the calibration of Meurer et al. (1999).
The UV slope, $\beta$, of the SED is estimated on the basis of the
$V-I$ color and the redshift with the relation
\begin{equation}
\beta = 3.23(V_{606} - I_{814})_{\rm AB} - 5.22 + 2.66z
- 0.545z^2 
\label{e:bphot2}
\end{equation}
and the absorption at 160nm $A_{1600}$ is obtained using the
conversion: 
\begin{equation}
A_{1600} = 4.43 + 1.99\beta.
\label{e:afit}
\end{equation}
The Calzetti extinction law (\cite{calzetti97}) has been adopted to
estimate the empirical $E_{B-V}$ (Column~9) from $A_{1600}$.
For a Salpeter IMF ($0.1 M_{\odot} < M < 100 M_{\odot}$) with constant SFR,
a galaxy with SFR $=1~ M_{\odot}$ yr $^{-1}$ produces 
$L(160$nm$)= 9.19 \times 10^{39}$ erg ~ s$^{-1}$ \AA$^{-1}$
(\cite{meurer99}).
The inferred star formation rates in the present sample range from 
a few tens to few hundred $M_{\odot}$ yr$^{-1}$. 

\begin{figure}
\includegraphics[width=90mm]{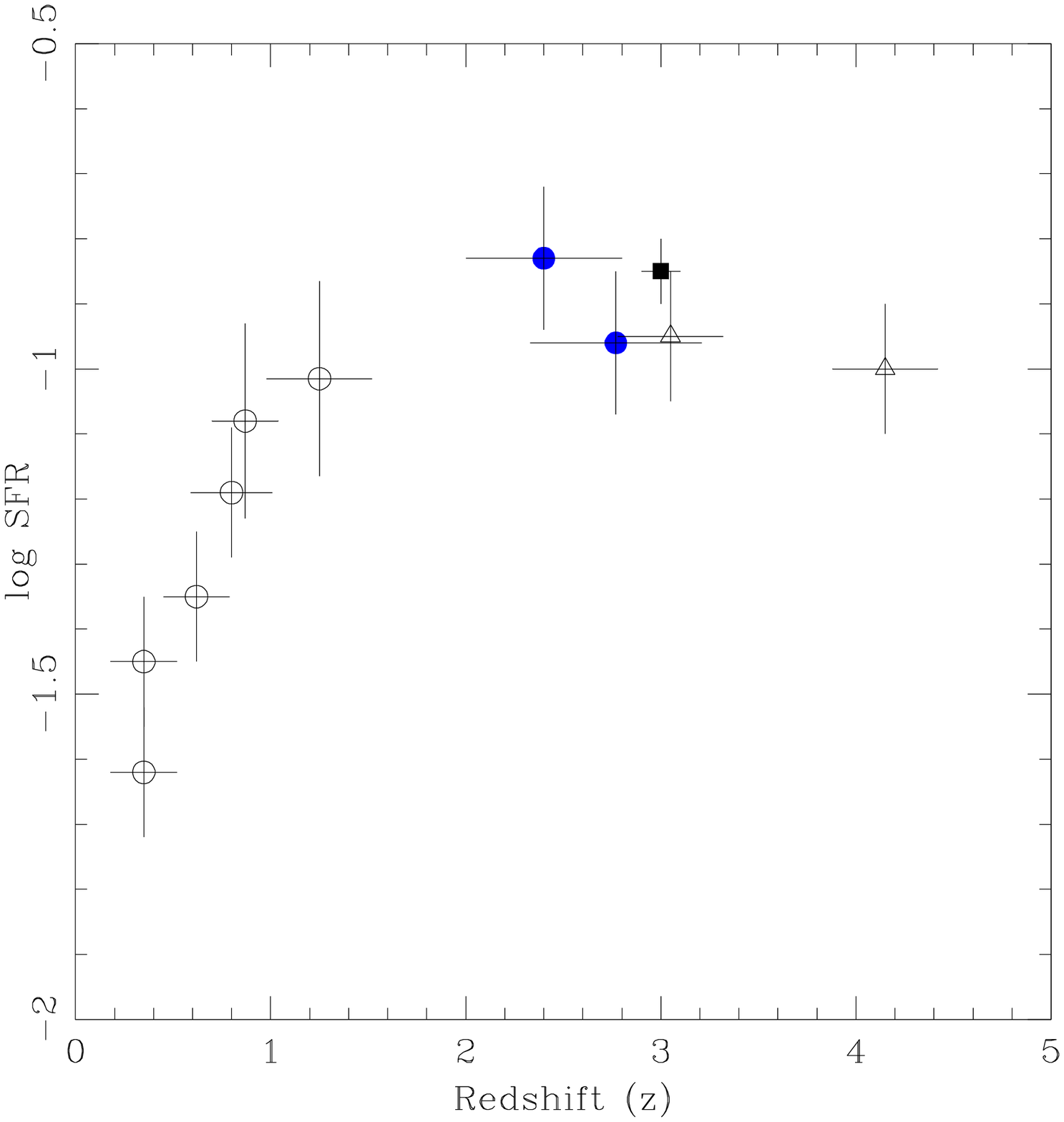}
\caption{Global star formation rate ($M_{\odot}$ yr$^{-1}$ Mpc$^{-3}$) as a function of redshift,
derived from complete spectroscopic surveys. Low redshift data ($z<2$) are taken from
\cite{lilly96} and \cite{wils02}. The higher redshift points (triangles) are from the $z=3$ and 
$z=4$ Lyman break galaxy samples of \cite{steid99}. The square symbol shows the X-ray estimate of 
the contribution to the global SFR from the $z\sim3$ LBGs (\cite{nandra02}). The filled circle 
at $z\sim2.4$ shows our estimate of the contribution to the global SFR, derived from the high-z sample
in the present work. The filled circle at $z\sim2.8$ is the estimate obtained in the HDF-N sample.}
\label{fig:globalSFR}
\end{figure}

By summing all the SFRs of Col. 4, we obtain an estimate for the global
star formation rate of the Universe of 0.056  $M_{\odot}$ yr$^{-1}$ Mpc$^{-3}$ at $<z> \simeq 2.4$.
The completeness limit of the present sample, $I_{AB}=24.25$ corresponds to a $M_{1700}=-20.47$=$M_{\star}$+0.37
(assuming a typical $V-I=0.25$ for our Lyman break galaxies),
where $M_{\star}=-20.84$ at 1700\AA$~$ is derived from \cite{poli01}.
Integrating over the whole luminosity range of the luminosity function a correction factor
of 2.7 is readily obtained, corresponding to an estimate for the universal SFR of 
0.15 $\pm$ 0.04 $M_{\odot}$ yr$^{-1}$ Mpc$^{-3}$. This is in good agreement with recent estimates in the literature
as shown in Fig.~\ref{fig:globalSFR}. The same procedure carried out on the HDF-N (\cite{cohen00}, \cite{soto99})
down to $I_{AB}=25.35$
provides an estimate for the universal SFR of 0.11 $\pm$ 0.03 $M_{\odot}$ yr$^{-1}$ Mpc$^{-3}$ at a median redshift of 2.8. 

\section{Individual notes}

An evident gravitational lensing effect is present in the HDF-S field.
The system consists of an arc at a radius of 0.9'' around an
elliptical galaxy (\cite{bark00}). 
We have obtained a spectrum of the lensing galaxy ($ID=1088$, $Nobj=37$),
which turns out to be an elliptical at redshift 0.577 (see also \cite{Denn01}). In this way the results
obtained by Barkana et al. assuming $z_{phot} \simeq 0.6$ for the lensing galaxy are 
confirmed.

We have also tried to obtain a redshift for the arc by comparing two spectra
taken at different position angles. In the first configuration the slits includes
both the galaxy and the arc, while in the second only the galaxy falls in the slit.
Unfortunately, better S/N spectra are needed to distinguish useful spectroscopic features
and obtain a reliable redshift of the arc.

Since the arc is not a U-dropout, we can put an upper limit to its redshift
of about 2.5, while a lower limit is provided by the redshift of the lensing galaxy, 0.577. 

In the HDF-S there is an interesting radio source identified and
discussed by Norris et al. (1999). This object (ID=373) is unusual in
that its radio-optical ratio is about 1000 times higher than that of
any galaxy known in the local Universe. It has also been
identified as an Extremely Red Object (ERO) with an
$(I-Ks)_{AB}=3.45$. In absence of strong emission lines the faintness
of the object ($I_{AB}=25.75$) has not allowed us to measure a
spectroscopic redshift. The absence of strong emission lines, however,
suggest that the photometric redshift estimated 
by Fontana et al. (2002), $z_{phot} \sim 1.7$,
is more likely than other, higher redshift possibilities
(\cite{norris99}).
%
%__________________________________________
\begin{acknowledgements}
We warmly thank  A. Arag\'on-Salamanca, C.Chiosi, A. Renzini, 
P. van der Werf and S. White
for stimulating the conception and the development of this work.
This study has been conducted with partial support by the TMR program
Formation and Evolution of Galaxies set up by the European Community
under the contract FMRX-CT96-0086.
This work was also partially supported by the ASI grants under the contract
number ARS-98-226 and ARS-96-176, by the research contract of the 
University of Padova ``The High redshift Universe: from HST and VLT to NGST''
and by the Research Training Network ``The Physics of the Intergalactic 
Medium'' set up by the European Community under the contract HPRN-C12000-00126 
RG29185. We also thank the referee for comments and suggestions that
greatly improved the paper.

\end{acknowledgements}

\begin{table}
\caption[]{Spectroscopic identifications in the HDF-S}
\begin{tabular}{lccccc}
\hline \hline
Nobj & $\alpha_{2000}$ & $\delta_{2000}$ & z & $I_{AB}$ & $ID$\\
\hline
1      &22:32:20.22 &-60:32:05.5  & 0.669 & -&- \\
2      &22:32:23.04 &-60:32:14.0  & 0.5125 & -& -\\
3      &22:32:27.29 &-60:32:28.2  & 2.055? &- &- \\
4      &22:32:30.81 &-60:32:47.7  & 0.579 & -& -\\
5      &22:32:32.88 &-60:35:33.8  & 2.256 & -& - \\
6      &22:32:33.40 &-60:36:59.7 & 0.3704 & -& - \\
7      &22:32:35.75 &-60:36:22.4 & 0.4727 & -& - \\
8      &22:32:36.40 &-60:35:18.2  & 0.3678 &- &- \\
9      &22:32:36.68 &-60:36:57.2 & 0.5153 & -& - \\
10     &22:32:36.78 &-60:36:06.4 & 2.641 & -& - \\  
11     &22:32:36.94 &-60:35:13.0  & 0.4209 &- &- \\
12     &22:32:37.34 &-60:31:57.6  & 1.214? & -&- \\
13     &22:32:38.37  &-60:36:49.6  & star & -&- \\
14     &22:32:39.00 & -60:36:32.0 & 0.7567 &- &- \\
15     &22:32:39.67 &-60:34:44.7  & 0.425 & -&- \\ 
16     &22:32:40.63 &-60:34:47.0  & 0.277 & -&-\\
17     &22:32:41.20 & -60:35:41.2 & 0.4138 & -& -\\
18     &22:32:42.15 & -60:37:14.5 & 0.3000 &- & -\\
19     &22:32:44.35 & -60:31:11.5 & 0.5146 & -& - \\
20     &22:32:44.91 & -60:30:55.5 & 0.5125 &- &- \\
21     &22:32:45.60 &-60:34:18.7 & 0.4594 &- &- \\
22     &22:32:45.68 & -60:36:55.5 & 0.5256 &- & -\\
23     &22:32:46.94  & -60:31:47.3 & 3.160 & - & 1527 \\
24     &22:32:47.80 & -60:32:19.6  & 2.077 & 24.22&1432 \\
25     &22:32:48.84 & -60:32:03.6 &  -  & 24.09 & 1314\\
26     &22:32:48.85 & -60:31:54.0 & 2.796 & 23.80& 1312\\
27     &22:32:49.09  & -60:32:05.9 & 2.162? & 23.75& 1284 \\ 
28     &22:32:49.18 &-60:32:26.0  & 2.789  & 23.28&1277 \\
29     &22:32:49.39 &-60:32:26.3  & 2.789  & 22.92&1242 \\
30     &22:32:50.00 & -60:33:45.1 & - & 23.63& 1178\\
31     &22:32:50.04 & -60:33:56.8 & 2.652 & 24.14 & 1174 \\ 
32     &22:32:50.47 & -60:30:36.1 & star & - &- \\
33     &22:32:50.51  &-60:31:03.0  & 0.5959 & -&- \\
34     &22:32:50.68  &-60:33:28.5 & 2.027 &23.32 & 1100\\ 
35     &22:32:50.70  &-60:33:25.9  & 1.1369 & 24.00& 1098 \\
36     &22:32:50.73  &-60:33:26.2 & 1.1369 &24.68 & 1095 \\
37     &22:32:50.90 & -60:32:43.0 & 0.5776& 20.90 & 1088 \\
38     &22:32:50.90 &-60:32:43.0 & - & 24.79$^\dag$& arc \\
39     &22:32:50.95 &-60:43:15.4 &   -  & 24.07& 1074 \\
40     &22:32:51.10 &-60:34:08.2 &   -  & 23.53& 1053 \\
41     &22:32:51.12 & -60:34:08.0 & 0.1798&- &- \\
42     &22:32:51.17 & -60:32:34.0 & - & 24.28 &1061 \\
43     &22:32:52.06 & -60:31:40.8 & 0.5116 & 22.60& 973\\
44     &22:32:53.47 & -60:35:24.8 & 3.5161&- &- \\ 
45     &22:32:53.85 &-60:32:13.1  & - &23.86 &771 \\
46     &22:32:53.92 & -60:33:13.5 & 0.3645 & 21.08& 758\\
47     &22:32:54.02 & -60:33:05.6 & 0.5804? & 22.74 & 746\\
48     &22:32:54.09 & -60:31:42.7 & 0.512 & 22.76& 751 \\
49        &22:32:54.80  &-60:31:13.9  & 0.546 & -&- \\
50     &22:32:54.81 & -60:31:13.9 & 0.512 &- &- \\
51     &22:32:54.82 & -60:31:31.2 & 0.513 & -&- \\
52     &22:32:55.36 & -60:34:35.5 & 2.643 &- &- \\
53       &22:32:55.65  &-60:31:16.8  & 0.5112 & -&- \\
\hline
\multicolumn{6}{l}
{$\dag$ Magnitude in Vega system, from Barkana et al. 2000.}\\
\label{tab:HDFS}
\end{tabular}
\end{table}
\setcounter{table}{1}
\begin{table}[h]
\caption[]{Spectroscopic identifications in the HDF-S}
\begin{tabular}{lccccc}
\hline \hline
Nobj & $\alpha_{2000}$ & $\delta_{2000}$ & z & $I_{AB}$ & $ID$\\
\hline
54       &22:32:56.20  &-60:31:07.9  & 0.5108 & -&- \\
55     &22:32:57.72 &-60:34:08.6 & Mstar &23.05  & 427 \\
56     &22:32:58.60 &-60:33:46.6  & -      & 25.75& 373 \\
57     &22:32:59.78 &-60:35:15.7 & Mstar &- &- \\
58     &22:33:03.21 & -60:33:29.7 & 2.025 & 23.25&115 \\ %%%%
59     &22:33:03.86  &-60:33:08.8  & 0.970  &22.98 &86 \\
60     &22:33:04.90 & -60:33:30.7 & 2.360?& 24.20& 28\\ %%%% 
61     &22:33:05.79  &-60:34:36.8  & 0.564  &- &- \\
62     &22:33:06.17 & -60:30:44.3 & Mstar &- &- \\
63     &22:33:06.19 & -60:30:56.2 & star &- &- \\
64     &22:33:09.53  &-60:34:36.2  & 1.160  &- &- \\
65     &22:33:12.65  &-60:33:50.1  &  Mstar &- &- \\

\hline
\end{tabular}
\end{table}

\begin{table}
\caption[]{Spectroscopic redshifts of objects in the WFPC2 HDF-S field.}
\begin{tabular}{lcccccc}
\hline \hline
ID & $\alpha_{2000}$ & $\delta_{2000}$ & z & $I_{AB}$ & ref\\
\hline
	  1527:  &  22:32:46.94 & -60:31:47.3&3.160 &  -&1 \\  
          1440:  &  22:32:47.57 & -60:34:08.6&0.560  & 21.22&2 \\ 
          1435  &  22:32:47.66 & -60:33:35.9&0.580  & 19.53&3 \\ 
          1432  &  22:32:47.80 & -60:32:19.6&2.077 &  24.22&1 \\        
          1312  &  22:32:48.85 & -60:31:54.0&2.796 &  23.80&1 \\        
          1309  &  22:32:48.88 & -60:32:16.1&0.580  & 22.96&4  \\ 
          1284  &  22:32:49.09 & -60:32:05.9&2.162? &  23.75&1 \\        
          1277  &  22:32:49.18 & -60:32:26.0&2.789 &  23.28&1 \\        
          1242  &  22:32:49.39 & -60:32:26.3&2.789 &  22.92&1 \\        
          1178  &  22:32:50.00 & -60:33:45.2&1.152 &  23.63&1 \\        
          1174  &  22:32:50.04 & -60:33:56.8&2.652 &  24.14&1 \\        
          1160  &  22:32:50.29 & -60:32:03.3&0.414  & 22.23&4  \\       
          1100  &  22:32:50.68 & -60:33:28.5&2.027 &  23.32&1 \\        
          1098  &  22:32:50.70 & -60:33:25.9&1.137 &  24.00&1 \\        
          1095  &  22:32:50.73 & -60:33:26.2&1.137 &  24.68&1 \\        
          1088  &  22:32:50.90 & -60:32:43.0&0.579  & 20.90&1  \\       
          1015  &  22:32:51.51 & -60:33:37.6&0.577  & 22.12&4  \\       
           973:  &  22:32:52.06 & -60:31:40.8&0.512 &  22.60&1\\
           948  &  22:32:52.14 & -60:33:59.6&0.564  & 21.79& 4 \\       
           936  &  22:32:52.24 & -60:34:02.8&0.511  & 22.11& 4\\       
           935  &  22:32:52.30 & -60:33:08.5&0.583  & 21.09& 4\\       
           853  &  22:32:53.03 & -60:33:28.5&1.270 &  22.60&2\\        
           820  &  22:32:53.33 & -60:32:39.3&1.114 &  22.46&4 \\        
           790  &  22:32:53.64 & -60:32:36.0&0.365  & 21.15&  4 \\       
           789  &  22:32:53.70 & -60:32:06.2&1.116 &  23.29&4 \\        
           775  &  22:32:53.74 & -60:33:37.6&0.565  & 21.48 & 4 \\       
           758  &  22:32:53.92 & -60:33:13.5&0.364  & 21.08&1 \\       
           746  &  22:32:54.02 & -60:33:05.6&0.580?  & 22.74&1 \\       
           743  &  22:32:54.05 & -60:32:51.7&0.515  & 22.14&4 \\       
           751:  &  22:32:54.09 & -60:31:42.7&0.512 &  22.76&1\\
           667  &  22:32:54.68 & -60:33:33.2&0.173  & 21.25 & 4 \\       
           592  &  22:32:55.72 & -60:32:11.5&0.673  & 21.59 & 4 \\       
           562  &  22:32:56.07 & -60:31:48.9&0.514  & 21.24 & 4 \\       
           443  &  22:32:57.54 & -60:33:06.1&0.583  & 20.46&3 \\       
           434:  &  22:32:57.75 & -60:32:33.0&0.517  & 22.00&4 \\      
           420:  &  22:32:57.99 & -60:32:34.3&0.760  & 21.34&2 \\      
           402  &  22:32:58.22 & -60:33:31.6&0.423  & 21.98& 4 \\      
           257  &  22:33:00.15 & -60:33:19.0&0.540  & 23.22&3 \\      
           254:  &  22:33:00.24 & -60:32:33.9&0.415  & 20.58 & 4 \\      
           182  &  22:33:01.77 & -60:34:13.5&1.230 &  22.43&2 \\       
           154  &  22:33:02.45 & -60:33:46.5&0.696  & 22.29&3 \\      
           141  &  22:33:02.76 & -60:33:22.1&0.465  & 20.31& 4\\      
           115  &  22:33:03.21 & -60:33:29.7&2.025 &  23.25&1 \\        
            98  &  22:33:03.57 & -60:33:41.7&0.340  & 20.07& 4 \\      
            86:  &  22:33:03.86 & -60:33:08.8&0.970  & 22.98&1 \\       
            28  &  22:33:04.90 & -60:33:30.7&2.360? &  24.20&1 \\        

\hline
\multicolumn{6}{l}
{1) this work.} \\
\multicolumn{6}{l}
{2) Franceschini et al. (in preparation).}\\
\multicolumn{6}{l}
{3) $http://www.aao.gov.au/hdfs/Redshifts$ }\\
\multicolumn{6}{l}
{4) Dennefeld et al. (in preparation).}\\

\label{tab:zspecALL}
\end{tabular}
\end{table}

\begin{table*}[h]
\centering
\caption{Properties of the Galaxies with $z>2$ in the HDF-S field.}
\begin{tabular}{rcccccccc}
\hline \hline
ID & $I(AB)$ & $zspec$ & $SFR(GIS)$  &
AGE & MASS & $E_{B-V}(GIS)$ &
$SFR(emp)$ & $E_{B-V}(emp)$
\\
&  & & $(M_{\odot}yr^{-1})$  &
(Gyr) & $[Log(\frac{M_{*}}{M_{\odot}})]$ & (mag) &
$(M_{\odot}yr^{-1})$ & (mag)
\\
\hline
\hline
%# id    MAG     Zspec  SFR  AGE    MASS        EBV    SFRcor  
1100 & 23.32 & 2.027 & 86 & 0.40 & 10.54 & 0.30 & 132 & 0.26 \\
 115 & 23.25 & 2.025 & 97 & 0.20 & 10.29 & 0.30 & 124 & 0.24 \\
1432 & 24.22 & 2.077 & 39 & 0.51 & 10.30 & 0.30 &  70 & 0.28 \\
1284 & 23.75 & 2.162 & 23 & 0.32 &  9.87 & 0.15 &  47 & 0.15 \\                       
  28 & 24.20 & 2.360 & 22 & 0.05 &  9.05 & 0.15 &  28 & 0.10 \\
1174 & 24.14 & 2.652 & 66 & 0.04 &  9.42 & 0.25 &  86 & 0.21 \\
1242 & 22.92 & 2.789 &211 & 0.01 &  9.34 & 0.20 & 188 & 0.16 \\
1277 & 23.28 & 2.789 & 60 & 0.20 & 10.08 & 0.15 & 124 & 0.14 \\
1312 & 23.80 & 2.796 & 52 & 0.51 & 10.42 & 0.20 &  95 & 0.17 \\
\hline
\hline
\label{tab:physics}
\end{tabular}
\end{table*}

\end{document}